%% file: main.tex
\documentclass[USenglish,oneside,twocolumn]{article}

\usepackage[utf8]{inputenc}%(only for the pdftex engine)
\usepackage[big]{dgruyter_NEW}

\usepackage{xcolor}
\usepackage{tabularx}
\usepackage{url}
\usepackage{graphicx}
\usepackage{csquotes}

\usepackage{tikz}

\renewcommand{\arraystretch}{1.4}
 
\DOI{10.56553/popets-2022-0117}

\cclogo{\includegraphics{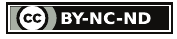}}
  
\begin{document}

  \author*[1]{Alexandra Kapp}

  \affil[1]{Hochschule für Technik und Wirtschaft Berlin (HTW), University of Applied Sciences, E-mail: 
  kapp@htw-berlin.de}
  \title{\huge Collection, usage and privacy of mobility data in the enterprise and public administrations}

  \runningtitle{Collection, usage and privacy of mobility data in the enterprise and public administrations}

  %\subtitle{...}

  \begin{abstract}
{Human mobility data is a crucial resource for urban mobility management, but it does not come without 
	personal reference. The implementation of security measures such as anonymization is thus needed to 
	protect individuals' privacy. Often, a trade-off arises as such techniques potentially decrease the utility 
	of the 
	data and limit its use. While much research on anonymization techniques exists, there is 
	little 
	information on the actual implementations by practitioners, especially outside the big tech context. 
	Within 
	our study, we conducted expert 
	interviews to gain insights into practices in the field. We categorize purposes, data sources, analysis, and 
	modeling tasks to provide a profound understanding of the context such data is used in. 
	We survey privacy-enhancing methods in use, which generally do not comply with state-of-the-art 
	standards of differential privacy. We provide groundwork for further research on practice-oriented 
	research by identifying privacy needs of practitioners and extracting relevant mobility characteristics for 
	future standardized evaluations of privacy-enhancing methods.}
\end{abstract}
%  \classification[PACS]{}
 % \communicated{...}
 % \dedication{...}

  \journalname{Proceedings on Privacy Enhancing Technologies}
\DOI{10.56553/popets-2022-0117}
  \startpage{440 }
  \received{2022-02-28}
  \revised{2022-06-15}
  \accepted{2022-06-16}

  \journalyear{}
  \journalvolume{2022}
  \journalissue{4}

\maketitle

\section{Introduction}
Nowadays smartphones are being used daily for a variety of functions - from mobile phoning through 
navigation to renting an e-scooter via an app. The usage of these applications produces data about the 
locations and movements of individuals, so-called \textit{human mobility data}, which can be a great 
resource to optimize 
services but also for a multitude of diverse tasks such as traffic planning 
\cite{naboulsi_large-scale_2016} or epidemiological research 
\cite{lai_measuring_2019}. 
As such data entails highly personal information, it falls under the European General Data Protection 
Regulation (GDPR) which restricts companies from freely using such collected data for arbitrary purposes. 
While the analysis of human mobility data offers great potential, it can be assumed that not all desirable 
use cases are implemented due to uncertainty regarding privacy regulations. A recent 
study~\cite{bitkom_research_ds-gvo_2020} concludes that 46\% of German companies 
refrain from innovations because of ambiguities in the interpretation of the GDPR. For example, 31\% 
claimed to not have implemented new technologies based on Big Data or Artificial Intelligence because of it 
and 41\% stated that they were unable to set up data pools or share data with business partners. 

Anonymization\footnote{\enquote{Anonymization} is a misleading term, as it suggests that data becomes 
fully 
	anonymous. Numerous examples of successful reidentification of individuals in \enquote{anonymized} 
	data 
	suggests 
	otherwise, e.g., \cite{ohm_broken_2009}, \cite{douriez_anonymizing_2016}, 
	\cite{xu_trajectory_2017}.} of data can be used as a measure to enhance customers' privacy 
and simplify data usage for companies, as GDPR principles no longer apply once data is considered 
anonymous (Recital 26 GDPR). Therefore, one option to make use of data more confidently is the 
implementation of privacy-enhancing methods that sufficiently guarantee privacy. However, 
anonymization of mobility data is a difficult task since people's movements follow predictable patterns 
\cite{gonzalez_understanding_2008} that allow easy re-identification. 
Individuals have successfully been re-identified from \enquote{anonymized} taxi data 
\cite{douriez_anonymizing_2016}, out of highly aggregated mobile phone data 
\cite{xu_trajectory_2017}, or the aggregated count of customers per station 
\cite{pyrgelis_what_2017}. This already illustrates that procedures that guarantee sufficient anonymization 
from a legal point of view are partly considered insecure within the privacy community. 
While big tech companies such as Google, Apple, or Microsoft put effort into adopting state-of-the-art 
privacy concepts 
like local differential privacy \cite{cormode_privacy_2018}, it is doubtful that these are widely 
used outside the big tech industry \cite{hopkins_machine_2021}.

Making mobility data available in a privacy-sensitive manner is a complex and multi-faceted problem. 
There is typically a trade-off between utilizing data and protecting privacy and the legal and technical 
assessment of the anonymity of the data may differ.
It is not trivial to gain in-depth insights on data practices in the 
field as companies rarely share detailed information on data usage. Even companies such as the cellular 
network operator Telefónica that claims to use sophisticated anonymization techniques 
\cite{telefonica_unser_nodate} do not share details about their methods.

We aim to understand which privacy-enhancing methods for human mobility data are already in use by 
practitioners and which privacy needs are still present. Thus, a profound understanding of real-life 
practices in the work with respective data is necessary, as the suitability of privacy methods depends on 
the context they are applied in. For example, if the goal is to release 
reports with aggregated statistics to third parties, one could add noise to the aggregates as a 
comparatively simple method that likely provides reliable results. On the other hand, training a 
next-location prediction algorithm requires fine-granular data input and therefore other appropriate 
privacy-enhancing methods are needed.

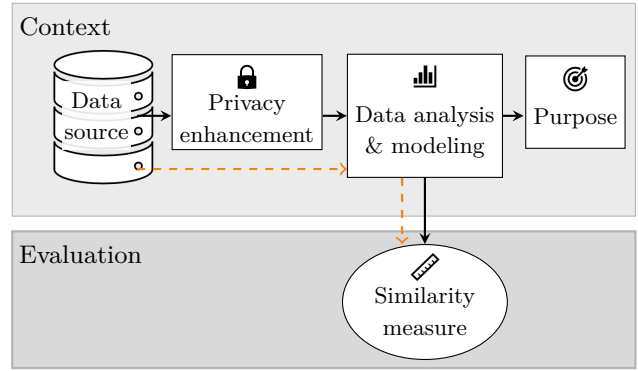
\begin{figure}[tb]
	\centering
	\input{overview}
	\caption{Schematic illustration of the context that privacy-enhanced methods are used in. Data sources 
		are 
		used (in a privacy enhanced manner) for data analysis or modeling tasks to achieve a certain purpose. 
		The evaluation of privacy enhancements is pursued by similarity measures which quantify the 
		difference 
		between analyses conducted with privacy-enhanced (black solid arrow) and raw (orange dashed 
		arrow) 
		data input. 
		Note, that for simplicity the use of privacy-enhanced methods is placed 
		between the data source and the data analysis, which is only one possible 
		set-up.}
	\label{fig:scheme}
\end{figure}

As shown in Figure \ref{fig:scheme}, we presume that mobility data serves as a data source to conduct 
analysis and modeling tasks which are means to acquire certain purposes.
For example, data from a public transport routing app 
(data source) is used to aggregate the number of routing 
queries for each hour of the day (data analysis) to optimize the operating hours of the 
public transport lines (purpose). 
With expert interviews, we aim to gain insights into these respective categories. In addition, we survey on 
privacy measures that are already in use.

Academic research evaluates their proposed privacy methods with \textit{similarity measures} which 
quantify the resemblance of analysis outputs with and without privacy enhancement. The more similar 
the two outputs remain the higher the utility of the privacy measure is rated. 
With this work, we further want to provide a link between real-life practices and academic research by 
extracting the core \textit{mobility characteristics} entailed in practitioners' use-cases so that relevant 
similarity measures 
can be identified. For example, analyzing the top 10 most used docking stations of shared bicycles entails 
at its core the same characteristic of interest as determining traffic volume on street segments: the spatial 
distribution of records. A standardized set of similarity measures that are matched to 
such characteristics would not only enable easier comparison between different privacy enhancing 
approaches but also 
simplify practitioners' assessment of suitable methods for acquired use cases. As similarity measures 
currently vary strongly within the literature it is difficult to compare different approaches, we 
hereby want to lay the groundwork for future standardization of such similarity measures.

In summary, our contribution consists of the following: 
(1) We provide a profound insight into real-life practices stated in expert interviews by employees from 
companies and public administrations in Germany working with human mobility data.
(2) We deduce core mobility characteristics as groundwork for the categorization and standardization of 
similarity measures.
(3) We identify privacy needs of practitioners.

This paper is organized as follows. In Section \ref{background}, we give an overview of related work on 
human mobility data. Section 
\ref{method} describes our methodology for the data collection, processing, and evaluation of the expert 
interviews. In Section \ref{findings} the evaluation of the qualitative data is presented. Section 
\ref{sec:implications} provides implications for privacy needs and similarity measures deduced from the 
interviews. Finally, the results are 
summarized and discussed in Section \ref{discussion}.

\section{Background}
\label{background}

Different \textbf{techniques to collect} human mobility data are well documented in the literature, see e.g., 
\cite{barbosa_human_2018, 10.1145/3485125, wang_urban_2019}. 
This includes data from surveys, mobile phone data in the format of call detail records, GPS tracking 
devices, usually smartphones that produce spatially and temporally fine granular data, and locations users 
post on social media. Some surveys also name WiFi positioning systems 
\cite{toch_analyzing_2019}. The overviews of data sources mostly focus on openly 
available data sets or such that have been used for academic research. One can easily imagine what kind 
of human mobility data companies could potentially raise with different 
techniques. Fiore et al.~\cite{fiore_privacy_2020} name five examples of sources for 
micro-trajectory data: Location-based services, like Google maps, record the GPS position while the 
app is running, cellular network operators collect call detail records, municipalities collect MAC addresses 
via Wi-Fi probe messages of nearby smartphones, car navigation systems record the GPS data of the 
navigation device, banks register the shops their customers pay at. While all these are valid examples of 
potentially used datasets, to the best of our knowledge there are no systematic overviews of mobility data 
actually used by companies and the ways this data is handled.

Data on human mobility is a highly desired resource for various \textbf{purposes}. For example epidemic 
spreading of diseases is being studied \cite{lai_measuring_2019}, just recently during the 
COVID-19 pandemic \cite{schlosser_covid-19_2020}. There is also growing research 
focusing on deep learning approaches to predict the next location of a person 
\cite{10.1145/3485125}, for instance, to predict locations of affected people during disasters 
like earthquakes \cite{lu_predictability_2012}. For an overview of various machine 
learning applications that human mobility data is used for, see \cite{toch_analyzing_2019}. 
Dedicated research also focuses on visual explorations of mobility data 
\cite{andrienko_visual_2013} and more and more interactive tools enable users to 
visualize large-scale fine granular mobility data\footnote{for example kepler.gl and unfolded.ai}. These 
examples are only a few of various use cases that build on 
human mobility data though this does not necessarily reflect applications in enterprise settings.

There is plenty of research dealing with \textbf{privacy-enhancing methods} for mobility data, for a 
detailed 
overview on methods for trajectory micro-data see Fiore et al.~\cite{fiore_privacy_2020}. To give a few 
examples: There are simple approaches like the reduction of granularity of coordinates 
\cite{gruteser_anonymous_2003} or reducing the sampling interval 
\cite{hoh_enhancing_2006}. More advanced methods aim to provide indistinguishability between 
individuals within a dataset, like  \textit{k}-anonymity \cite{bennati_privacy_2020}, or provide 
uninformativeness with the guarantee of 
differential privacy, e.g., \cite{gursoy_differentially_2019, acs_case_2014, 
	to_differentially_2016, ranise_practical_2016}. 

While privacy researchers consider differential privacy as the de-facto standard, there 
is little information on the adoption of such methods in the field. Garfinkel et 
al.~\cite{garfinkel_issues_2018} point out that the deployment of differential privacy comes with 
challenges 
and requires skilled staff. Calacci et al.~\cite{calacci_tradeoff_2019} 
state that risk and utility are often evaluated without context which is vital for a proper assessment. They 
analyze the public and market utility as well as the risks associated with different levels of granularity of 
mobility data, 
thereby only considering coarsening and aggregating as privacy enhancement, which they say is still 
most commonly used in practice. De Montjoye et al.~\cite{de_montjoye_privacy-conscientious_2018} also 
criticize the insufficient implementation of privacy measures for mobile phone data and propose four 
different approaches for practical implementations in real-life scenarios. While both, Calacci et al. and de 
Montjoye et al. assume reasonable scenarios, we aim to collect empirical data on the context of mobility 
data usage.

Privacy-enhancing methods reduce the information content and thus there is a common perception of an 
associated reduction of utility of the data. This is true for many use cases, for example, when a public 
transport 
company wants to analyze the typical distance their customers are willing to walk to a stop, the utility is 
likely
decreased when the exact locations are obfuscated with noise or aggregated to larger grid cells. Other use 
cases are less impacted by such measures, for example, those that are based on highly aggregated data 
such as the evaluation of customer numbers over time of a new 
bike-sharing system. Thus, it is vital to understand the analysis purposes and methods that are 
applied 
in practice to evaluate the trade-off between utility and 
privacy when privacy-enhancing methods are applied. Similarity measures are commonly used in research 
to quantify 
the utility, 
though there do not exist any standard measures for privacy-enhancing methods applied to mobility 
data~\cite{fiore_privacy_2020}.
In addition to a (potential) impact on the utility, other effects of privacy-enhancing methods also ought to 
be considered in practice, e.g., research about medical data shows that users are more willing to share 
data 
when they have trust that their privacy is preserved~\cite{kalkman_patients_2022}.

\section{Methodology}
\label{method}
In July and August 2021 we conducted a total of 13 semi-structured expert interviews that lasted on 
average 
about one hour, with a range between 30 minutes and 1.5 hours. 
The interviews covered questions on mobility data sources, including their origin, structure, and personal 
reference.
Further questions dealt with data analysis and modeling techniques, their purposes as well as the impact 
they have on the companies' actions. Additionally, we asked about analyses planned for the future, those 
that have not been conducted due to (legal) restrictions or obstacles, and data protection and 
anonymization practices.
Questions were asked about how long data is stored, in which format, and whether 
anonymization techniques are applied.
Questions on data security, the legal basis, and user communication were 
included, however, they are not further evaluated within this work. See Appendix \ref{apx:interviewGuide} 
for a 
the full interview guide.

\subsection{Participant recruitment and moderation}
All interviewees were employees in leading positions of German organizations working with human mobility 
data. One organization was represented with two interviewees from different departments, thus resulting 
in twelve different organizations of the following types: public administrations, public transport companies, 
a mobility platform (part of 
a public 
transport company), a mobility service provider, an automobile manufacturer, a location-based 
service app, a sensor company providing sensors for people counts, and market research 
companies. The location-based service app was still in the state of a startup and did not work with any real 
customer data yet, but the participant could report on planned data usage. Also, one public administration 
only recently started with a dedicated team to work with human mobility data and one person from a public 
transport company reported mostly from their current build-up process. The rest of the interviewees 
had multiple years of experience with mobility data within their field and company. All participants were in 
the positions of founders, CEOs, team or project leads of relevant divisions. See Table \ref{tab:participants} 
for an overview of all participants which also introduces the participants' IDs which will be used in Section 
\ref{findings}.

\begin{table}[]
	\small\sf\centering
	\caption{Overview of interview participants.}
	\label{tab:participants}
	\begin{tabular}{lll}
		\textbf{ID} & \textbf{Organization type} & \textbf{Job title}            \\
		\midrule
		P1          & public administration      & mobility manager              \\
		P2          & public administration      & manager in traffic mgmt. \\
		P3          & public administration      & head of the data science \\
		P4          & public transport company   & team lead AI systems          \\
		P5          & public transport company   & product owner analytics       \\
		P6          & public transport company   & team lead offer planning      \\
		P7          & mobility platform          & project lead                  \\
		P8          & mobility service provider  & managing director \\
		P9          & automobile manufacturer    & head of analytics             \\
		P10         & location-based service app & CEO                           \\
		P11         & sensor company             & head of technical division    \\
		P12         & market research company    & CEO                           \\
		P13         & market research company    & managing director       \\      
	\end{tabular}
\end{table}

We recruited participants through contacts of our research group network and by sending email invitations 
to relevant company representatives (see Appendix \ref{apx:email} for the email invitation text). We used 
purposeful sampling \cite{palinkas_purposeful_2015}, a method where a variety of relevant cases is sought 
to be included by specifically targeting a selection of such differing subjects, i.e., a variety of different 
types of organizations working with mobility data.

All interviews were held in German and conducted remotely using a video conferencing tool. The interviews 
were recorded and transcribed with the aid of transcription software. The automatically created 
transcripts were proofread and corrected by the interviewer.

\subsection{Analysis and coding}
The qualitative content analysis with an inductive approach~\cite{Kuckartz2019} was conducted to analyze 
the interview transcripts. All questions within the interview guide were constructed to fit into one of the 
following groups (see Appendix \ref{apx:interviewGuide}) which match the scheme in Figure 
\ref{fig:scheme}: 
(1) purposes, 
(2) data sources, 
(3) data analysis and modeling (initially: methods), 
and (4) privacy (the code on user communication an legal was not used for the evaluation). 
All relevant parts of the transcripts were extracted into a table format and 
categorized into one of the four codes.
Each group was then evaluated separately. During the iterative coding phase each 
transcript chunk was coded on a fine-granular level first, then the codes were grouped into broader 
categories.  The coding process was conducted by the interviewer herself and the coding iterations were 
discussed with and reviewed by one further person; the revisions served as the base for further 
refinements.
For the analysis of purposes, we aimed to find common motivations and themes which are printed in bold 
type in Section \ref{purposes}. 
Data sources were categorized based on technical similarities and are listed in Table \ref{tab:sources}.
The major distinction between applications was the computation of statistical aggregations (see all final 
codes in Table \ref{tab:analysis}) and the 
application of mathematical models (see all final codes in Table \ref{tab:modeling}).
Themes for the motivation of anonymization were derived (see final themes printed in bold type in Section 
\ref{privacy}); applied and planned privacy-enhancing methods were collected (see final codes in Table 
\ref{tab:privacy}).

\subsection{Research ethics and anonymization}
During the recruitment, participants received information about the purpose of the study, and an informed 
consent document (see Appendix \ref{apx:consentForm}) was signed before the interview. The interviewer 
also informed 
participants verbally 
about the purpose of the research and the audio recording. Additionally, they were guaranteed  
precautious handling of the shared information which does not allow to identify the person or 
company. To further stress that no sensitive 
information about the person or company would be revealed, this information was again repeated within 
the verbal introduction at the beginning of the interview (see Appendix \ref{apx:interviewGuide}). After the 
transcription, all 
audio files were 
deleted and the names of participants and their employers were removed from the transcripts. The 
transcripts were then stored in an encrypted manner. See Appendix \ref{apx:studyProc} for the full study 
procedure with respect to ethical considerations.

\subsection{Limitations}
Participants were only recruited from companies based in Germany. Thus, results can only be applied to a 
limited amount of companies in other countries. As GDPR was of special interest, the results can most 
easily be transferred to other EU countries.
We have made our best effort to recruit diverse organizations, though we cannot claim to have included all 
types of organizations working with human mobility data.
Our main research focus is directed at companies working on urban mobility topics, therefore we recruited 
our participants accordingly. Still, we are aware that companies from other contexts also work with 
such data, for example, location-based service apps like fitness apps, restaurant 
recommendations, or dating apps, to only name a few.

\section{Findings}\label{findings}

Participants are referred to by there ID P1-P13 as assigned in Table \ref{tab:participants}.

\subsection{Purposes for data collection and analyses}
\label{purposes}

The primary purpose for collecting personal data is usually the operation of a service. As P8 stated, they 
cannot provide route suggestions if they don't know where the 
customer wants to go to. 
However, this evaluation focuses on determining themes for data analysis and modeling purposes beyond 
the operation of applications. 

Several experts mentioned that data is used for \textbf{demand-driven offers} to customers. For example, 
the P8 (mobility service provider) said they position their vehicles close to the predicted demand and plan 
their 
fleet size accordingly. They also optimize routing algorithms for ride-hailing applications based on 
customer data. P6 (public transport company) stated that they not only do mid- to long-term offer 
planning 
but also adapt their schedules within a few weeks 
to better suit the changing needs during the COVID-19 pandemic. They added that ticket options and 
pricing are also 
part of the long-term offer planning that relies on customer mobility data. \textbf{Quality management} was 
mentioned by P8, for example by comparing 
the actual and predicted waiting time of taxi customers. P9 (automotive manufacturer)
named the 
need for data for \textbf{autonomous driving}.

Data is not only used internally but it is also used to \textbf{provide information to customers}: P6 said they 
use historic data to 
predict future passenger loads and display such information in their routing application. P9 stated that the 
display of real-time traffic can help car drivers to avoid traffic 
jams.

Insights from aggregated data have been stated to be used for \textbf{marketing}. Personalized 
advertising 
has only 
been named by one expert as a potential option that is not intended to be pursued.

Various experts named \textbf{reports of aggregated statistics for monitoring of KPIs, 
	internal knowledge, and strategic 
	planning}. Not only to obtain new insights but also to verify gut feelings, as one expert said: 
\enquote{Every [manager] knows [...] the customer behavior very well. [...] [They] 
	have a feeling, an experience, but it's better to really see it in black and white} [translated from German].

P6 and P3 (both public administration) stated to use data for \textbf{city and 
	traffic planning}, e.g., to plan bicycle infrastructure. They also envisioned other fields of application for 
the 
future, 
e.g., to compute emissions produced within a city by the transport sector or to create plans for 
emergency or catastrophe situations.

The \textbf{provision of data to third parties} was mentioned in various forms: P3 mentioned 
their efforts of providing as much data as possible as open data for transparent policymaking. Some 
interviewees said they are obliged 
to provide data to other parties, for example, public transport companies need to report aggregated 
statistics 
to the public administration. One expert also claimed they are considering to potentially sell anonymized 
data in the future. All data 
the public administration has access to can potentially 
be subject to parliamentary inquiries. Two experts also reported on the use of data for evidence in court.

\subsection{Data sources and responsibilities}
There are two types of data that practitioners work with: data collected by themselves and external data, 
such as open data, bought data, or data provided by contractual partners.
The question of origin plays a major role in terms of who needs to have technical and legal competencies  
on the protection of privacy: if data is 
gathered through external sources the responsibility is seen with the providing entity. The provider needs 
to have 
the competencies of applying adequate privacy measures while the party receiving data is (at most) 
interested in the high-level information on whether the data is GDPR compliant.

As P11 (sensor company) stated: \enquote{[...] We have such a certificate for our solution. Well, what we 
attach 
	to the 
	tender and show: Okay, look at our solution, it works. But it is also compliant with the General Data 
	Protection Regulation. [...]
	Otherwise, we wouldn't be able to offer the solution in the market in Germany or anywhere else or in 
	Europe [...]} [translated from German].
While P13 (market research company) said: \enquote{There I notice the tendency that the industry clients, 
so to 
	speak, they like to play down the data protection requirements a bit in order to get this data more 
	quickly, so as not to 
	make things so difficult for themselves} [translated from German].

Table \ref{tab:sources} summarizes the categorized data sources named by the interviewees which are 
already being used or desired to be used in the future.

\begin{table*}[tb]
	\small\sf\centering
	\caption{Data sources by type, providing entity, user, and available format as stated by the experts.}
	\label{tab:sources}
	\begin{tabularx}{\textwidth}{lXXX}
		\textbf{Type} & \textbf{Provision} & \textbf{User \newline (among   interviewees)} & 
		\textbf{Available format} \\ 
		\midrule
		\textbf{surveys} & third party research   institutes & public administrations (P1-P3), \newline public 
		transport 
		companies (P4-P6) & 
		aggregated and anonymized data \\ 
		\textbf{stationary sensor data} & sensor companies maintain sensors and preprocess data & 
		public   
		administration (P3), \newline public transport company (P6) & preprocessed and anonymized data \\
		\textbf{routing queries} & app operator & mobility 
		service   provider (P8), \newline public transport companies (P5, P6), \newline mobility platform (P7)&
		data users are also app controllers, thus, access to data in any (legally permitted) way  \\ 
		\textbf{transaction data }& app operator & mobility 
		service   provider (P8), \newline mobility platform (P7) & data users are also app controllers, thus, 
		access 
		to data 
		in 
		any 
		(legally permitted) way  \\
		\textbf{GPS tracking} & controller of   tracking device applications (e.g., smartphone apps, vehicles 
		equipped 
		with   GPS trackers) & mobility service  provider (P8), \newline mobility platform (P7), \newline public 
		transport 
		company (P6), \newline public 
		administration (P3), \newline
		market research company (P12), \newline automobile manufacturer (P9), \newline location-based 
		service app (P10) & 
		heterogeneous: 
		If data users are also app controllers they have access to data in any (legally permitted) way. 
		Third parties can only access aggregated and anonymized data.  
		\\ 
		\textbf{mobile phone data} & cellular network  providers & public administration (P3), \newline 
     	public 
		transport company (P6) & aggregated and   anonymized origin-destination matrices \\ 
	\end{tabularx}\\[5pt]
\end{table*}

Large-scale household \textbf{surveys} are a traditional mobility data source that all 
interviewed public administrations and public transport companies stated to rely on. 
The experts mentioned additional custom studies on smaller scales that are commissioned by public 
administrations or companies. They all agreed that surveys are commonly conducted by third-party 
research facilities who are responsible for the data privacy concept while only aggregated and anonymized 
data is made available to third parties.

Unlike other forms of mobility data where the person carries the tracking device, \textbf{stationary 
	sensors} are 
positioned statically and people passing the sensor are registered. As the sensor provider (P11)
explained, there are specialized companies that install and run sensors, provide software, make the sensor 
signals human-readable, and take care of anonymization measures if the data contains personally 
identifiable 
attributes. Technical variations that interviewees stated include pressure sensors within the 
road surface (e.g., to measure traffic 
volume), infrared sensors (e.g., to count entering and exiting public transport passengers), camera-based 
sensors (e.g., to count people within a room) or sensors based on WiFi technology that allows the tracking 
of MAC addresses of mobile devices across multiple sensors. Only camera and WiFi-based sensors were 
seen as potentially critical in terms of privacy.

Routing applications provide information on the optimal route and potential alternatives, based on a 
provided start and end location and time. \textbf{Routing queries} made within such apps precede many 
actual trips and can be considered a proxy to mobility data. P5 and P6 from public transport 
companies reported that they 
collect such data with their own routing applications and use it for analytical purposes, e.g., for passenger 
load forecast. As app operators, they both stated to have raw data access which is 
restricted by technical and organizational measures.
Usually, query data is not stored with any user identification which limits the personal references. Though, 
as people tend to query routes to sensitive and personal locations, like home or work, privacy concerns 
could be raised, 
as P6 also mentioned.

Apps that allow the booking of mobility services produce mobility-related \textbf{transaction data}. 
Transaction 
data includes the exact start and destination location as well as time, price, and user information. This data 
is primarily needed to handle the booking transaction with the payment, as P8 said, but is also used for 
aggregated statistics. P2 from a public administration 
reported obtaining aggregated statistics of such transaction data from a 
partnering bike-sharing provider. Unlike routing queries, transaction data includes information on actually 
performed trips with precise time 
and place information and a linked user record. It is therefore highly personal information.

\textbf{GPS tracking} data is diverse due to different types of devices and applications. Some experts 
stated to 
collect GPS data themselves, either via their smartphone applications or GPS-equipped vehicles. For 
example, the market research company (P12) offers an app that constantly tracks participants during the 
collection phase of a study. GPS data is also acquired from third parties: P3 (public 
administration) reported that they considered using aggregated data about street-level speed and  
traffic volume that had been collected with an app for cyclists and P6 (public transport) reported about 
planning a market 
research study which includes GPS tracking.

\textbf{Mobile phone data} is collected by cellular network providers. P6 explained that they buy such data 
from a service provider who gets anonymized data 
from cellular network providers. The service provider then processes them into usable formats such as 
origin-destination matrices and redistributes them.

\subsection{Data analysis and modeling techniques}
\label{sec:analysis}

Methods stated by the experts can mainly be assigned to one of two groups: statistical 
aggregations and mathematical models. While statistical aggregations provide descriptive analytics of the 
data, mathematical models, i.e., machine learning models and traffic models, allow tasks such as 
classification, prediction, or simulation. 

All stated \textbf{statistical aggregations} are presented in Table \ref{tab:analysis}.
They are grouped according to shared underlying characteristics 
which are generic attributes of mobility data 
independent from the specific context.

\begin{table*}[tb]
	\small\sf\centering
	\caption{Statistical aggregations used by the interviewed experts and their underlying mobility 
	characteristics.}
	\label{tab:analysis}
	\begin{tabularx}{\textwidth}{XX}
		\textbf{Statistical aggregations} & \textbf{mobility characteristic} \\
		\midrule
		trip counts, customer counts, returning customers & record counts \\
		total passengers over time, bike rentals over time & temporal distribution of records \\
		people count per location / sensor, top 20 shared mobility stations, sold tickets at a station, 
		passengers entering and exiting a station, number of transits per station, traffic volume, 
		occupancy  rate in a place / public transport line, real-time traffic information & spatial distribution of 
		records \\
		\textit{all aggregations for spatial distributions disaggregated by certain time windows} & spatial and 
		temporal   
		distribution of records \\
		mobility demand by OD relations, round trips of shared bikes (i.e., same start and end station) & 
		distribution 
		of OD counts\\
		relation of public transport share compared to other modes by OD relation & modal split per OD pair \\
		average trip lengths (evaluated in research studies)  & trip length  \\
		dedicated analysis of trip chains (evaluated in research studies) & travel patterns \\
		daily driven distances (car), temporal changes in daily distances (e.g., to see trends during COVID-19 
		pandemic 
		or 
		holidays) & daily range \\
		modal split (evaluated in research studies) & modal split \\
		trips conducted with multiple traffic modes (e.g., bike \& ride) (evaluated in research studies) &  
		inter-modality of trips\\
		proportion of people who use more than one traffic mode (evaluated in research studies) & 
		multi-modality of people \\
		average speed per street segment (bicycle and car) & speed \\
		waiting times at traffic lights, customer time spent in stores & time allocation \\
		customer groups (e.g., x\% of customers visiting store A also visit store B) & correlation between visits 
		of different locations \\
	\end{tabularx}\\[5pt]
\end{table*}

On the highest level, mostly all experts aggregate data to record counts like the total number of trips or 
customers. 
On a more fine-granular level, different experts are interested in spatial (and temporal) distributions that 
quantify people at 
certain locations (and times), e.g., public transport companies (P6) are interested in the number of 
customers 
entering and exiting stations (at different times of day) and the number of tickets sold per station, while 
the mobility service provider (P8) wants to know where (and when) their services are mostly used.

Origin-destination matrices are used by public transport companies (P6, P13) to gain insights into 
mobility demand, 
further disaggregated by modes of transport they reveal shortcomings in public transport infrastructure. 
According to P13, surveys and market research studies evaluate average trip lengths, trip chains, the 
modal split, as well 
as the share of inter-modal trips (i.e., multiple modes of transport are mixed within a trip) and multi-modal 
people  (i.e., a person uses multiple traffic modes for different trips). 
P3 (public administration) stated to be interested in data on speed and traffic volumes of bicycles per 
road segment. Additionally, they were interested in waiting times at 
traffic lights. 
P11 (sensor company) stated that clients are interested in identifying customer groups based on their 
visited 
locations. Additional performance indicators such as the number of unique customers per location, the 
number of returning customers per location, and time spent at different stores are determined.

Next to statistical aggregations, experts (plan to) use data for different \textbf{mathematical models}, listed 
in Table \ref{tab:modeling}. Due to the nature of such models, underlying characteristics cannot 
be determined in the same manner as before.
P8 (mobility service provider) uses demand prediction models and optimizes 
routings of ride-hailing services to optimally group users. 
P11 (sensor provider) explores the prediction of people counts, though 
they do not see any demand for such features among their customers.
P12 (market research company) reported to use classifiers that detect the mode based on continuous 
smartphone GPS 
tracking and further sensor data. They also 
experimented with activity recognition algorithms which are supposed to recognize the purpose of a visit, 
such as \enquote{at home} or \enquote{waiting}. 
Public administrations (P2, P3) and public transport companies (P6) reported using traffic models, 
commonly 
4-step-traffic-models\footnote{The 4-step-traffic-model is a travel demand model to forecast traffic 
	following four steps: (1) trip generation, (2) trip distribution, (3) mode choice, (4) route choice 
	\cite{mcnally_four_2007}.}, which take a variety of data sources as input such as population density, 
modal split and origin-destination matrices to simulate different scenarios and forecast traffic.  
According to P2 and P3, agent-based models are also in the planning which require user trajectories of an 
entire day to properly 
take trip chains into account.
Predicting the next probable location of a user (next-location prediction) was in a proof of concept stage at 
the location-based service app (P10). They also planned on implementing an algorithm 
to cluster customers' mobility behavior. 

\begin{table}[tb]
	\small\sf\centering
	\caption{Mathematical models in use and in planning by interviewed experts.}
	\label{tab:modeling}
	\begin{tabular}{l|l}
		\textbf{in use} & \textbf{in planning} \\
		\midrule
		4-step traffic models & next-location-prediction \\
		occupancy prediction & activity recognition \\
		mode detection & clustering of mobility patterns \\
		routing optimization & agent-based models \\
		demand prediction & \\
	\end{tabular}
\end{table}

\subsection{Privacy}
\label{privacy}
We found major differences regarding the \textbf{engagement of the interviewees with privacy 
	measures}.
We hypothesize that there is a difference between participants' organizations that \textbf{collect data 
themselves and those that obtain them from third parties}. 
For example, experts from public administrations (almost) 
exclusively work with third-party human mobility data, therefore they did not report any need of 
implementing anonymization 
methods themselves. Still, privacy is an important topic for them, as data protection authorities strictly 
check any personal data that is used by public administrations.

All interviewees applying anonymization methods to their data named one of the 
two \textbf{reasons}:  
(1) For purposes \textbf{outside of the scope the user consented to}.
(2) To make the data \textbf{available to third parties}.
Different experts reported that they \textbf{struggle to pursue all their use cases due to GDPR}. 
They said, that personal data 
cannot be used for any arbitrary purpose, even if it might serve the customers' interests.
Therefore, anonymization techniques can help to remove the personal reference and enable 
the use for additional analyses. As one interviewee explained: 
\enquote{There is a source layer [...][with] 
	GPS in full resolution and whatnot. This is normally not usable at all for analysts like me and after 
	processing [and anonymizing] it is moved to the secondary assets. The primary assets for the primary 
	use case are then 
	deleted} [translated from German].

Interviewees with a business model based on \textbf{providing data to third parties}, such as market 
research companies or the sensor provider, have a \textbf{high interest in applying privacy measures} 
as \textbf{compliance with GDPR is a major criterion to acquire clients}. Accordingly, they seemed to have 
the highest expertise in privacy-enhancing methods. Table \ref{tab:privacy} shows an overview of 
privacy-enhancing methods that were stated by the experts.

\begin{table*}[tb]
	\small\sf\centering
	\caption{Privacy-enhancing methods and their contexts within the experts' organizations.}
	\label{tab:privacy}
	\begin{tabularx}{\textwidth}{XX}
		\textbf{Privacy measure / guarantee} & \textbf{Context} \\
		\midrule
		\textbf{removal of personal attributes} & storage of recorded GPS locations without any customer 
		information\\
		\textbf{pseudonymization} & sensor company pseudonymizes MAC-address recorded with WiFi 
		sensors \\
		\textbf{aggregation} & (1) aggregate data from surveys/studies as reports \newline (2) 
		dashboard 
		with 
		aggregated bike   sharing data provided to public administration \newline
		(3) internal knowledge   sharing of 
		insights based on statistics \\
		\textbf{indistinguishability} & (1) market 
		research   
		company (P12)  provides origin-destination information only for connections above a certain 
		threshold 
		\newline (2) mobile phone 
		data is 
		provided only for connections above a certain threshold \newline 
		(3) 
		State Office of Statistics provides spatially aggregated data only for cell counts > 5 \newline (4) 
		 automobile manufacturer (P9)  only includes POIs in analyses that exceed a  certain user count 
		 threshold \\
		\textbf{coarsening} & (1) heatmaps instead of maps with single points are used to visualize study 
		results 
		\newline 
		(2) P9 rounds coordinates to three decimal places for the analysis of POIs \\
		\textbf{cropping of trajectories} & (1) P9 crops trajectories for the analysis of 
		frequently 
		used road 
		segments \newline (2) P3 names cropping as a known best practice and role model 
		for potential future release of anonymized open data \\
		\textbf{noise} &P12 uses different anonymization techniques based on 
		the 
		analyses: 
		adding of noise is mentioned as one option \\
		\textbf{synthetization} & P12 investigated synthetization options but evaluated 
		the 
		utility as not   sufficient for their sample sizes \\
		\textbf{differential privacy} &P12 tests differential privacy methods to exempt 
		data from 
		being strictly bound to study purposes \\
		\textbf{de-centralized data processing} & P12 envisions to run certain 
		algorithms 
		(e.g., mode detection) directly   on user devices in the future; for privacy reasons but also for 
		faster   processing capabilities \\
	\end{tabularx}\\[5pt]
\end{table*}

P4, P9 and P11 stated to \textbf{remove personal information} that is not needed for analyses, such as 
name, 
phone number, or MAC address. Some experts claimed to remove the user identification entirely while 
others 
still retained the link between different user records but used \textbf{pseudonymization} methods on the 
user ID. 
Data \textbf{aggregation} is not only a method for analytical purposes but also a measure of 
anonymization, as stated by P2, P11 and P12.
P2, P6, P12 and P9 reported that data is restricted such that locations counts of spatially aggregated 
data 
need to surpass a certain threshold to be accessible, thereby providing \textbf{indistinguishability}.
%with \textbf{\textit{k}-anonymity} restrictions, 
Two of them received such restricted data from third parties while two implemented such measures 
themselves. 
P9 reported working with a \textbf{reduction of granularity of coordinates and timestamps 
	(coarsening)} for start 
and 
end locations. P12 explained that map views would only show heatmaps instead of exact points 
as a visual implementation of reduced granularity. \textbf{Cropping of the beginning and end of a 
	trajectory} for 
fine granular GPS trajectory data was reported by P9 and P12 also stated to 
\textbf{add noise to the data}.

Advanced methods like \textbf{synthesizing data}\footnote{On the basis of raw data a new synthetic data 
	set is 
	created that, depending on the used algorithm, maintains certain statistical distributions from the original 
	dataset.}, methods that implement \textbf{differential privacy} and \textbf{de-centralized 
	data 
	processing} have 
only 
been 
named by one expert (P12) as methods that are being tested within the organization for potential future 
usage.

\section{Practical implications}
\label{sec:implications}

\subsection{Privacy needs of practitioners}

Based on the interviews, we can identify different privacy needs of practitioners.

A common scenario is the compilation of pre-defined aggregated statistics. While the experts did not 
see privacy needs in addition to aggregations, privacy research suggests otherwise 
\cite{xu_trajectory_2017, pyrgelis_what_2017}.
Since many analyses in different contexts are based on similar characteristics (see 
Table~\ref{tab:analysis}) a 
set of proven privacy-enhancing methods and tools for standard analyses could be helpful.

However, not all useful analyses are known in advance. Data is used in exploratory scenarios and new use 
cases 
arise. As one expert said: \enquote{We repeatedly have questions [that could be answered with the survey 
	data]. 
	But for data protection reasons it was promised that the data will be deleted at the end of last 
	year} [translated from German].
Data release is another relevant scenario: Data used for decision processes of public administrations is 
desired to be published as 
open data. Also, agent-based traffic models are desired to be used but they need single user trajectories 
as input which are usually not shared by data providers. Data synthetization techniques could be a viable 
privacy enhancement where data remains in the original format and can be used for arbitrary purposes and 
without time restrictions. Though it should be noted that synthetization 
techniques maintain only certain statistical properties depending on the specific algorithm. There is an 
increasing amount of research on synthetization of mobility data, but these methods are far from 
established and practically proven. They need to be evaluated carefully and if applied, limitations of the 
utility need to be well communicated.

The operation of applications based on machine learning models like mode detection, 
next-location-prediction, or activity recognition need fine-granular input data which cannot be obfuscated 
or aggregated in advance. Differentially private adaptations of machine learning algorithms can be used to 
limit the impact of single users onto the model and thereby the potential privacy breaches. Also, 
de-centralized approaches, like federated learning, could be considered to prevent centralized storage of 
personal data.

The lack of expertise to assess which anonymization techniques are sufficient causes insecurity and 
lengthy processes. As one participant said: \enquote{[...] it is not so easy to find 
	expertise that covers both technical know-how on data level and can serve the legal perspective as well. 
	[...]
	If someone says I want to do this, but the data must be anonymized for that, we have to involve a lot of 
	other people who tell us how to do it and who can also somehow give the okay for it to be really legally 
	secure} [translated from German]. Concrete recommendations for action would provide guidance for 
faster processes and implementations.

Finally, it should be noted that there is a need for easy-to-use tools that can also be implemented by 
organizations that do not have the resources or expertise for employees with dedicated skills on privacy 
methods. The more accessible such methods are, the likelier the gap between research and practice 
will shrink.

\subsection{Similarity measures}
Utility losses due to privacy-enhancing methods are quantified with similarity measures, as shown in 
Figure~\ref{fig:scheme}.\footnote{There is 
	no standard name for such measures, different publications also use the following terms:
	measure (or metric) of utility, evaluation, resemblance, dissimilarity, quality, accuracy, information loss, 
	or utility loss. } They determine 
how much a characteristic, e.g., the spatial distribution, of privacy-enhanced data resembles the 
output generated with the raw data. As researchers evaluate their proposed privacy methods on varying 
similarity 
measures results are hard to compare amongst them. Similarity 
measures might address different characteristics or 
even different nuances of a characteristic, for example, the spatial distribution can be captured by 
quantifying how many of the top 50 locations are identified correctly or by comparing the 
distribution of location visits with the Jensen-Shannon divergence. Depending on the use case, 
different characteristics need to be maintained by privacy-enhancing methods, thus different similarity 
measures are relevant.

With Section~\ref{sec:analysis} we want to provide guidance on relevant characteristics for a future 
categorization and standardization of such similarity measures. While the definition of characteristics is 
fairly straightforward for statistical aggregations, suitable measures for mathematical models are more 
difficult to evaluate. Either a more profound understanding of such models is needed to derive respective 
characteristics or privacy methods need to be evaluated directly on accuracy measures of downstream 
tasks, 
e.g., correctly detected traffic modes by a mode detection algorithm with and without 
privacy enhancement.

\subsection{Recommendations}

In summary, we can derive the following recommendations:

\begin{itemize}
	\item To provide practitioners with guidance and clarity on the use of state-of-the-art privacy-enhancing 
	methods for mobility data, an easily accessible framework could be useful which compiles practical 
	real-world use cases and suggests adequate privacy methods.
	The handout for companies published by Germany’s digital association Bitkom on	
	\enquote{Anonymization and 
	pseudonymization of data for machine learning projects} is an illustrative example 
	of such a publication about a related topic \cite{aichroth_anonymisierung_2020}.
	\item A provision of easy-to-use tools for privacy enhancing methods will enable organizations without 
	the expertise and resources to implement state-of-the-art methods. Such tools could provide a 
	compiled report of typical mobility analyses or the generation of synthetic data.
	A project like the \textit{Synthetic Data Volt (SDV)} \cite{7796926} which is an overall system 
	for synthetic data models, benchmarks, and metrics could be extended for mobility data or serve as an 
	example for a similar approach.
	\item A set of standardized similarity measures and downstream tasks would facilitate the comparison of 
	different privacy enhancing methods and enable practitioners to choose the most suitable method for 
	their use case. The SDV package includes model agnostic metrics which could again serve as an 
	example or be extended with mobility data specific metrics.
	
	\item GDPR certificates for privacy-enhancing technologies could accelerate the processes within 
	organizations and provide security for decision makers.
\end{itemize}

\section{Discussion} \label{discussion}

Movement data undoubtedly holds great potential for commercial as well as scientific analyses. However, 
the highly individual patterns in the data, which make them so interesting, mean that anonymization is 
hardly possible without utility losses. The high legal attention to the processing of such data leads to 
frequently encountered challenges in practice which motivated us to take a detailed look at the data used 
in organizations and the analysis and anonymization methods that are being applied. We 
conducted 13 interviews with employees of German companies and public administrations working with 
human mobility data.
Even though many assumptions are made concerning the practical use of such data, to the best of 
our knowledge, this is the first systematic study to evaluate such 
sources, usage, and privacy measures in enterprises. We grouped and listed our results to 
provide an overview of real-world practices with such data and identified different scenarios of privacy 
needs of practitioners. Thereby, these insights can be used as a basis 
for future research on practice-oriented privacy-enhancing techniques and tools 
that help to close the gap between research and practice. 

The interview evaluation shows a detailed 
breakdown of data sources in use, including their origin and 
available formats. This information can guide future privacy research regarding target groups and use 
cases 
for proposed methods.

Compliance with GDPR is a major concern stated by many experts. Thereby, legal requirements are almost 
exclusively the origin of instating privacy measures. This is in accordance with 
Beringer et al.'s~\cite{beringer_usable_2021} findings who see a need for a regulatory framework for 
usable privacy and security and conclude that business interests are mainly directed at collecting as 
much data as possible. 
Though, 
much uncertainty remains about possible techniques, 
their implications on utility, and tools to implement those in practice. 
Expertise of anonymization 
techniques strongly varies among organizations and largely 
depends on whether data is gathered and used by themselves, provided to third parties, or if it is only 
received by data providers. 
While academic research has accepted differential privacy as the de-facto standard, it is not 
yet implemented in practice, if known at all. One expert also 
stated that they neither have the time nor the expertise to implement advanced methods. Providing 
easy-to-use tools to simplify the implementation of privacy-enhancing methods is thus a necessary step to 
increase the usage of such methods. Especially companies that have no dedicated business case of 
providing anonymized data usually lack such resources.

To increase the accessibility of methods that state-of-the-art research suggests, the utility for the actual 
data analysis purposes of practitioners needs to be ensured. Therefore, we see the need for a diverse 
palette of standardized similarity measures that cover different kinds of use cases. Proposed 
privacy-enhancing methods use varying similarity measures concerning different mobility 
characteristics. This makes the comparison and interpretation of the utility across different methods 
burdensome. We hope that our research provides a more comprehensive overview of the practical context 
of mobility data use cases and relevant mobility characteristics that help to develop a set of diverse 
similarity measures reflecting actual practitioners' needs.

\subsection*{Acknowledgements}
This work is part of the FreeMove project.
I acknowledge the financial support by the Federal Ministry
of Education and Research of Germany in the framework of the FreeMove project.  

I hereby thank Helena Mihaljević for the constructive feedback and the other project members for their 
valuable input.

\bibliographystyle{IEEEtran}
\bibliography{references.bib}

\appendix
\section*{Appendix}
\addcontentsline{toc}{section}{Appendices}
\renewcommand{\thesubsection}{\Alph{subsection}}

\subsection{Interview guide [translated from German]}
\label{apx:interviewGuide}
%\includepdf[pages=-,scale=.9]{appendix/interviewGuide}

\subsubsection*{Code assignment of questions}

\colorbox{cyan}{\textit{purposes}}
\newline
\colorbox{green}{\textit{data sources}}
\newline
\colorbox{yellow}{\textit{methods (renamed to: data analysis and modeling)}}
\newline
\colorbox{lightgray}{\textit{privacy}}
\newline
\newline
initial codes that were not used for further evaluation:
\colorbox{pink}{\textit{User communication and legal}}

\subsubsection*{Welcome}

\begin{itemize}
\item  \textit{Give a introduction of the research project.}
\item \textit{State objective of the interview:} \enquote{The interviews are a first step within our research 
project 
to 
capture the status quo of privacy of mobility data in practice: what data is available in the first place, how 
is 
it stored and analyzed. Therefore, in this interview, I would like to learn more from you about the three 
topics: Data collection, data use, and data storage. I will guide you through the interview based on 
various 
questions about these blocks, but this does not have to follow strict protocol - I welcome input that you 
find relevant beyond the questions.
In our project, data protection plays an important role - an often difficult and definitely sensitive topic. I 
would therefore like to emphasize again in advance that we do not want to imply any lack of measures 
with 
the questions or put you under any pressure to justify them. All questions that go into detail about data 
protection measures are solely intended to gain a better understanding of current practices so that we 
can 
bring our research closer to reality. No results will be published that could be construed negatively 
towards 
your company in any way. The results will of course be anonymized, i.e. no names of companies will be 
mentioned.
The same applies to the use of the data: we want to understand on a general level for which purposes 
data 
is needed. No information will be published about your specific use cases that could reveal potential 
business secrets.}
\item \textit{Ensure consent form has been signed.}
\item  \textit{Confirm verbally that the consent to audio record the interview has been given.}
\end{itemize}

\textit{- - - - - - - Start audio recording - - - - - - -}

\subsubsection*{Interview questions}
\textit{Examples in italic type writing can be used by the interviewer to clarify the question.}
\newline
\newline
\textbf{General}

\begin{itemize}
\item What is the product / service of your company?
\item How many employees are there in your company?
\item What is your position in the company?
\end{itemize}

\textbf{Data collection: What personal mobility data do you collect?}
\begin{itemize}
\item \colorbox{green}{\parbox{7.5cm}{Which personal mobility data do you collect yourself as a 
company?}}
\item \colorbox{green}{\parbox{7.5cm}{Which data do you purchase or get from third parties?}}
\item \textit{If there are multiple data sets:}
\begin{itemize}
\item Which of these data sets is the most relevant (the most challenging from a privacy perspective) to 
your work / is used the most? \newline \textit{(Focus on this data set for the rest of the questions)}
\end{itemize}
\item \colorbox{green}{What does the data look like in detail?}
\begin{itemize}
\item What geolocation technology is used to collect the data? \textit{(e.g., GPS, CDR, WiFi sensors)}
\item How temporally and spatially granular is the data?
\item Is there additional information about the collected locations? \textit{(e.g., semantic information about 
the 
locations, 
such as home, workplace, or restaurant)}
\item How long is the average duration of a trajectory?
\end{itemize}
\item \colorbox{lightgray}{Personal reference of the data}
\begin{itemize}
\item About which persons is data collected? \textit{(e.g., all customers, app users, people passing a 
sensor, 
...)}.
\item Over what period of time is mobility data available about a person? \textit{(e.g., anonymization after x 
days? 
New user ID every x days?)}
\item What other data is known about the user?
\textit{(e.g., demographic data, place of residence, purchase information, subscriptions / contracts)}
\end{itemize}
\end{itemize}

\textbf{Data use: How will the data be used to gain insights for your purposes?}

\begin{itemize}
\item \colorbox{cyan}{For what purposes is the data used?} \newline
\textit{(Question to get started on data use: reporting, optimizing pricing models, advertising, etc.? First 
ask in general terms, then ask in more detail for specific analyses that 
are used.)}
\item  \colorbox{yellow}{\parbox{7.5cm}{What types of analyses or modeling are performed? With what 
goal?}}
\textit{(Depending on the answer, ask further in detail.)}
\begin{itemize}
\item Aggregate statistics
\item Detailed analysis of individual areas or users
\item Models, for prediction or classification
\end{itemize}
\item \colorbox{yellow}{At what frequency are the analyses conducted? }
\textit{(e.g., regular reports, real-time, one-time analyses)}
\item \colorbox{yellow}{How have the analyses evolved over time?}
\textit{(e.g., have more been added steadily / become more complex, have different ones been tried and 
discarded)}
\item \colorbox{yellow}{\parbox{7.5cm}{What role does exploration of data, without specific prior targeting, 
play in your 
work?}}
\item \colorbox{yellow}{Which explorations are carried out here?} \newline
\textit{Ask for a specific example: }what did the last exploration look like? What data, what analyses?
\item \colorbox{yellow}{\parbox{7.5cm}{Is there additional data that you combine with yours?}}
\textit{(e.g., purchased data, open data)}
\begin{itemize}
\item If so, which ones and how?
\end{itemize}
\item \colorbox{cyan}{\parbox{7.5cm}{What impact do the insights from the data have on your actions?}}
\textit{(e.g., positioning of mobility hubs, fare design, personalized advertising)}
\item \colorbox{yellow}{\parbox{7.5cm}{What further analysis or modeling is planned for the future?}}
\begin{itemize}
\item \colorbox{yellow}{\parbox{7.5cm}{What insights do you expect to gain from these analyses?}}
\item \colorbox{cyan}{\parbox{7.5cm}{What would be the potential impact of the findings?}}
\end{itemize}
\item \colorbox{yellow}{\parbox{7.5cm}{What further analyses or modeling would you do, assuming there 
were no 
hurdles?}} \textit{(e.g., 
amount of data, 
legal 
restrictions, computing capacity, or similar)}
\begin{itemize}
\item \colorbox{yellow}{\parbox{7.5cm}{What insights would you hope to gain from these analyses?}}
\item \colorbox{cyan}{What would be the implications?}
\item What hurdles exist to these analyses not being conducted?
\end{itemize}
\item \colorbox{lightgray}{\parbox{7.5cm}{Are there privacy measures being applied to the analysis?}}
\textit{(e.g., limit on queries, only certain queries, synthetic data generation, k-anonymity)}
\begin{itemize}
\item If so, by whom were these initiated?
\end{itemize}
\item \colorbox{lightgray}{\parbox{7.5cm}{What technical or legal constraints do you have on data use?}}
\item \colorbox{lightgray}{\parbox{7.5cm}{Are there any analyses or modeling that you have not done 
before due to privacy 
concerns? Which 
ones?}}
\end{itemize}

\textbf{Data storage: How is the data stored?}

\begin{itemize}
\item  \colorbox{lightgray}{How long is the data stored?}
\item  \colorbox{lightgray}{In which format is the data stored?} \newline
\textit{(e.g., database, single files)}
\item  \colorbox{lightgray}{Is the data being stored anonymously?}
\begin{itemize}
\item If yes, how?
\end{itemize}
\item  \colorbox{lightgray}{Who has access to the data?} \newline
\textit{(e.g., individuals, specific departments, the whole company)}
\item  \colorbox{lightgray}{How is this access documented and controlled?}
\item  \colorbox{lightgray}{\parbox{7.5cm}{Is the data passed on to third party data service providers?}}
\begin{itemize}
\item If so, in what form?
\end{itemize}
\item  \colorbox{lightgray}{Are there restrictions on access?} \newline
\textit{(e.g., are only certain queries possible? Is access to raw data possible?)}
\item  \colorbox{lightgray}{\parbox{7.5cm}{Are there data security measures that are taken in data 
storage?}}
\item  \colorbox{lightgray}{\parbox{7.5cm}{What other technical or legal restrictions do you enforce 
regarding data 
storage?}}
\end{itemize}

\textbf{User communication}
\begin{itemize}
\item  \colorbox{pink}{\parbox{7.5cm}{Are individuals informed about data collection or processing?}}
\begin{itemize}
\item If yes, how?
\end{itemize}
\item  \colorbox{pink}{\parbox{7.5cm}{On what legal basis is the data collected or processed?}}
\textit{(e.g., consent, contract, legitimate interests, legal basis)}
\end{itemize}

\subsection{Recruitment email [translated from German]}
\label{apx:email}

%\includepdf[pages=-]{recruitement\_email.pdf}

Dear NAME,

Within the framework of the BMBF-funded research project freeMove, we are working on a data 
protection-compliant use of personal mobility data. As an employee of COMPANY NAME, we cordially 
invite you to actively participate in our research project in the form of an expert interview. 

With the help of these interviews, we would like to gain a better understanding of the use of personal 
mobility data in practice. Accordingly, we would like to learn more from you about your daily work at 
COMPANY NAME. This knowledge will feed into the transdisciplinary research on privacy-compliant 
processing of mobility data. The aim of the research project is to develop practical and legally compliant 
recommendations for action that simplify work with personal mobility data and make it faster and more 
transparent.
The content of the interviews will be used for research purposes and will only be published after strict 
anonymization. 

CONTACT PERSON NAME is your contact person for scheduling an interview. 

DETAILS TO SCHEDULE A MEETING

More information about the research project can be found on our website www.freemove.space and in the 
attached PDF document. If you have general questions about the project process, goals, and initial project 
results, you can contact CONTACT PERSON EMAIL.
If you are unable to participate in the interview yourself, we would also be pleased if you could forward your 
questions to your colleagues. 

With kind regards
The freeMove Team

\newpage
\subsection{Text informed consent form [translated from German]}
\label{apx:consentForm}

%\includepdf[scale=0.9, offset = 30 -40, pages=-]{appendix/consentForm}

Research project: FreeMove
\newline
Performing institution: 	HTW Berlin
\newline
Interviewer:	 		Alexandra Kapp
\newline
Interviewee: xxx
\newline
Interview date:		xx.xx.2021

The BMBF-funded transdisciplinary project FreeMove explores privacy-friendly collection and analysis of 
mobility data. The aim of the project is to develop recommendations for action for the handling of personal 
mobility data.

As part of the scientific research project, the Department of Computer Science at the Hochschule für 
Technik Berlin (HTW Berlin) will conduct expert interviews with employees from administration and 
business. The purpose of the interview is to gain a sound understanding of the real-world handling and use 
of personal mobility data.

Personal data is processed, such as the name and employer of the interviewee, and other concrete 
information that could result from the interview because it is revealed by the interviewees. 
To facilitate the use of the study results and to verify or post-correct the notes written down by the 
interviewer, the interviews are recorded. In this process, the voice of the interviewee will be stored for the 
duration of the transcription process, but will be deleted no later than December 31, 2021.  

The transcription will be be supported by transcription software called 'Trint'. In this process, data may be 
transferred to the UK as 'Trint' is based in the United Kingdom (UK). Should data be transferred to the UK, 
this will be done on the basis of the European Commission's adequacy decision of 28 June 2021, which 
recognizes the UK as a third country with an adequate level of protection. Further information can be found 
in the privacy policy of 'Trint'. This is available at https://trint.com/privacy-policy.
The scientific analysis of the interview is carried out exclusively by the staff of the FreeMove research 
project. All employees who have access to the interview texts are obliged to maintain data secrecy.

All results will be published exclusively anonymously and without any possible conclusions about individual 
companies, organizations or persons.

Under the above-mentioned conditions, I agree to participate in the interview as part of the FreeMove 
scientific research project and consent to the recording, transcription, anonymization and analysis for the 
above-mentioned purpose. I also agree that my data may be processed using the software 'Trint' to 
facilitate the transcription process and that data may be transferred to the UK.
My participation in the interview and my hereby given consent to the processing of my personal data are 
voluntary. 

I am entitled at any time to request HTW Berlin to provide me with comprehensive information about the 
data stored about me.
I may at any time request HTW Berlin to correct, delete, block and transfer individual personal data, as well 
as to restrict processing. 
In addition, I can exercise my right to object at any time without giving reasons and modify or completely 
revoke the granted declaration of consent with effect for the future. For this purpose, an e-mail to 
Alexandra Kapp alexandra.kapp@htw-berlin.de is sufficient. I will not suffer any disadvantages as a result 
of refusal or revocation.

I hereby confirm that I have been informed in detail about the aim and the course of the research project 
and about my rights.

DATE, SIGNATURE INTERVIEWER

DATE, SIGNATURE INTERVIEWEE

\newpage
\subsection{Study procedure with respect to ethical considerations}
\label{apx:studyProc}

%\includepdf[pages=-]{research\_protocol.pdf}

\begin{enumerate}
\item Recruitment: email with detailed information about research objective
\item Interview
\begin{itemize}
\item Get signed informed consent form which informs about research objective, audio recording, 
transcription software, anonymization, and analysis of the interview
\item Provide verbal information about research objective and preservation of anonymity at the beginning 
of the interview
\item Get additional verbal consent on audio recording
\item Start audio recording and interview
\end{itemize}
\item Transcription
\begin{itemize}
\item Transcribe interviews with transcription software (explicitly stated in consent form)
\item Proof-reading by interviewer
\end{itemize}
\item Deletion of audio recordings
\item Anonymization: Removal of participants names and company names from transcripts
\item Data storage
\begin{itemize}
\item Storage of printed consent forms in a secured location of the research institution separated from 
transcripts
\item Encrypted storage of transcripts
\end{itemize}
\item Data evaluation
\end{enumerate}
\end{document}

%% file: overview.tex
\usetikzlibrary{shapes.geometric}

\tikzstyle{process} = [rectangle, minimum width=1.3cm, minimum height=1.35cm, text centered, 
scale=0.9, 
draw=black, fill=white]
\tikzstyle{line} = [thick,-,>=stealth]
\tikzstyle{arrow} = [thick,->,>=stealth]
\def\bottom#1#2{\hbox{\vbox to #1{\vfill\hbox{#2}}}}

\begin{tikzpicture}[node distance=1cm and 1cm]

	\draw[rectangle, fill=lightgray, opacity=0.3] (-1.2cm,1.3cm) rectangle (7cm,-1.5cm);
	\draw[line, fill=gray, opacity=0.3] (-1.2cm,-1.7cm) rectangle (7cm,-3.5cm);
	
	\node (context) [yshift=1cm, xshift=-0.5cm] {Context};
	\node(eval)[yshift=-2cm, xshift=-0.3cm] {Evaluation};
	\node (proDSimg) [yshift=-.2cm]
	{\includegraphics[width=2.2cm]{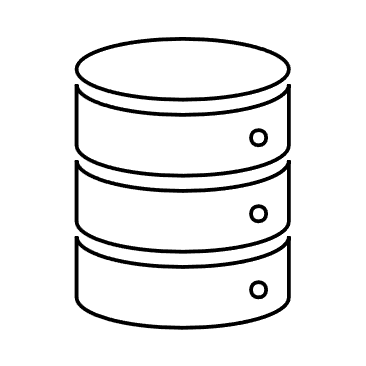}};
	\node (proDSbox) [rectangle, scale=0.9, fill=white, opacity=0.9, xshift=-0.1cm, yshift=-.2cm, minimum 
	height =1cm, minimum width = 1cm]{} ;
	\node (proDS) [rectangle, scale=0.9,xshift=-0.1cm, 	yshift=-.2cm, align=center] {Data \\ source};
	\node(proPEarrowbox) [rectangle, scale=0.9, right of=proDS, xshift=1.2 cm, align=center, minimum 
	width=2.2cm] {};
	\node (proPE) [process,right of=proDS, xshift=1.2 cm,yshift=0.2cm, align=center] {\bottom{0.7 
	cm}Privacy \\ 
	enhancement};
	\node at ([yshift=-0.3cm]proPE.north)
		{\includegraphics[width=4mm]{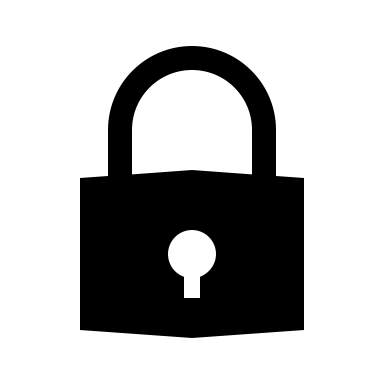}};
	\node(proDM)[process,right of=proPE, xshift=1.6 cm, yshift=-0.2cm, minimum height=1.8cm, 
	align=center] {\bottom{0.7cm} Data analysis \\ \& modeling};
	\node at ([yshift=-0.3cm]proDM.north)
{\includegraphics[width=4mm]{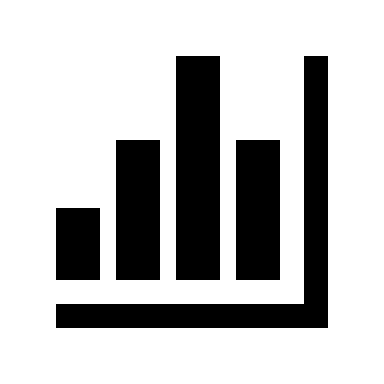}};
	\node(proParrowbox) [rectangle, scale=0.9, right of=proDM, xshift=1.2 cm, align=center, minimum 
width=1.5cm] {};
	\node(proP)[process,right of=proDM, xshift=1.2 cm, yshift=.2cm] {\bottom{0.7 cm}Purpose};
		\node at ([yshift=-0.3cm]proP.north)
	{\includegraphics[width=4mm]{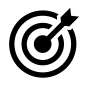}};
	\node(proSM)[ellipse, fill=white, draw=black, below of=proDM, yshift=-1.45cm,scale=0.9, align=center] 
	{\bottom{0.5
	cm}Similarity \\ 
	measure};
			\node at ([yshift=-0.3cm]proSM.north)
	{\includegraphics[width=4mm]{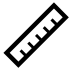}};
	
	\draw [arrow, yshift=0.2cm] (proDS) -- (proPEarrowbox);
	\draw [arrow] (proPEarrowbox) -- (proDM);
	\draw [arrow] (proDM) -- (proParrowbox);
	\draw [arrow] (proDM) -- (proSM);
	
	\begin{scope}[transform canvas={yshift=-0.7cm}]
		\draw [dashed,->, orange, thick]  (proDS) -- (proDM);
	\end{scope}
	
	\begin{scope}[transform canvas={xshift=-0.3cm}]
		\draw [dashed,->, orange, thick]  (proDM) -- (proSM);
	\end{scope}

\end{tikzpicture}